\documentclass[twocolumn,showpacs,superscriptaddress,floatfix]{revtex4}
\usepackage{graphicx}
\usepackage{bm}

\begin{document}
\title{Polarization transitions in Quantum Dot Quantum Well Arrays}
\author{Ethan Brown}\altaffiliation[Currently at ]{Physics Department, University
of Illinois at Urbana-Champaign.}
\affiliation{Homer L. Dodge Dept. of Phys. and Ast., University of Oklahoma, Norman,OK 73019}
\author{Kieran  Mullen,}
\affiliation{Homer L. Dodge Dept. of Phys. and Ast., University of Oklahoma, Norman,OK 73019}
\date{\today}
\begin{abstract}

With the improvement in fabrication techniques 
it is now possible to produce atom-like
semiconductor structures with unique electronic properties.
This makes possible periodic arrays of nano-structures
in which the Coulomb interaction, polarizability, and tunneling
may all be varied. 
We study the collective properties of  2D
arrays and 3D face centered cubic lattices  of singly-charged nano-spherical shells, sometimes called
`quantum-dot quantum wells' or `core-shell quantum dots.'
   We find that for square arrays,
the classical groundstate is an Ising anti-ferroelectret (AFE),  while
the quantum groundstate undergoes a transition from a uniform state
to an AFE.  The triangular lattice, in contrast, displays properties
characteristic of frustration.    
Three dimensional face-centered cubic lattices polarize in planes, with
each layer alternating in direction.   We discuss the possible
experimental signals of these transitions.

\end{abstract}
\maketitle

\section{Introduction}

\label{sec:intro}

With improvements in the colloidal fabrication of structures
on the micro- and nanometer scale, new areas of study have been
made possible.  In particular, researchers are now able to produce
atom-like electronic devices with their own unique spectra and shell
structures \cite{ph,pengdots}.  
The distinct characteristics
of such devices may be controlled through fabrication, in turn
producing structures with unique electronic properties.  

Periodic arrays of atoms have been
studied extensively for nearly a century.  However, we may now examine periodic arrays
of nano-structures in which the tunneling, polarizability,
and Coulomb interaction may all be varied, thus giving us the power to
control an electronic wavefunction to a degree without an atomic analogue.
Small multi-dot systems show 
coherent phenomena in experiments\cite{multidot}  while disordered arrays of colloidal quantum dots
display variable long-range hopping.\cite{hop} It should soon be possible
to examine collective electronic excitations of periodic nanostructures.
With this in mind, we choose a nano-structure which will confine
the electron on the nanoscale in such a way as to bring about novel
characteristics.  

In this paper we consider electrons confined to spherical shells,
sometimes called ``quantum dot quantum wells''  (QDQW) or ``core-shell
quantum dots.'' QDQWs  are heterostructures where layers of
different semi-conducting materials alternate in a single nano-crystal.
One well characterized example
is the CdS/CdSe/CdS system,  
i.e. a CdS core and outer shell with a CdSe inner shell
\cite{b}.
In this case each layer is on the order of 1-2 nm and is separated by a large
electronic band gap, thus allowing the investigation of quantum
confinement in a geometry in which electrons occupy the surface of
a sphere.  CdS/CdSe/CdS QDQW's have also been found to have an
electron-g factor which varies as a function of quantum well width
as well as a transverse spin lifetime of several nano-seconds at
temperatures approaching room temperature  \cite{b}.

We choose QDQW's for two reasons.  
First, because they have  
already be fabricated and are well-characterized,\cite{b} 
and second, because their excluded core structure
yields properties not found in regular quantum dot systems.  
Previous research\cite{rings}   has found that in a 1D array of singly-charged
nano-rings, a quantum phase transition ($T=0$) occurs which allows the
system to spontaneously break symmetry into anti-ferroeletric (AFE)
alignment.  Studies of the 2D ring
problem revealed another phase transition from random orientation
to a striped AFE state, analogous to unitary transition of the 2D
XY model.  The focus of this paper is to show an analogous
transition occurs in an ordered array of QDQW's.  

In the section below we describe our model, and then our methods for
analyzing it in the classical and quantum mechanical limits.  
We then present results for the 2D square lattice, the 2D triangular
lattice and the 3D face-centered cubic (FCC) lattice.  
Both the 2D square lattice and 3D face-centered
cubic lattice systems yield a
phase transition from random (classical) or uniform
(quantum mechanical) distribution to an anti-ferroelectric state.

\section{The Model}

\label{sec:model}

In our
model, each QDQW is treated as a infinitesimally thick 
spherical shell of radius $R$
with an electric charge on the surface. Charging might be achieved either
by doping,\cite{ddots}  tunneling from a backgate
 or
be adjusting an electrochemical potential.\cite{hop}  The radial degree of freedom can 
be neglected when the gap between radial excitations is large compared to that
of orbital eigenstates. 
We then consider periodic arrays of singly-charged QDQW's
in either  two or three 
dimensions and define a distance, $D$, between 
neighbors.
In 2D we choose $D$ as the lattice constant so that $D>2R$.  
In a 3D face-centered cubic lattice $D$ is half the distance
between vertices of the FCC cells so that the separation between
nearest neighbors is given by $\sqrt{2}D$.  We then consider the
electrostatic interaction solely between nearest neighbors.  Though
the Coulomb repulsion acts over large distances, we assume that
there is sufficient screening so that only nearest neighbor
interactions are significant.  Furthermore, the insulating
shell around each QDQW allows us to ignore electron tunneling between
dots. 


\subsection{Classical Analysis}

\label{sec:classical}

Before solving a quantum mechanical problem it is often helpful to
look at the classical case, which is usually easier to solve and provides an
guide to the quantum ground state. Classically the electron is a point charge constrained to the
surface of a sphere.  Because of this constraint, it has a dipole moment (with respect to the shell center) 
of constant magnitude
but random direction.  Adjacent moments interact leading to the polarization instability, as
we will show below.

There is only one energy scale
in the classical problem, the Coulomb energy $E_c=e^2/D$. The electrostatic
energy of the  array is given by
\begin{eqnarray}\label{eq:potential}
U=\sum_{i, j \in n.n.
\atop i\neq j  }^{N}{e^2\over |{\vec r}_i(\theta_i,\phi_i)-{\vec r}_{j}(\theta_{j},\phi_{j})|}
U={1\over 2} \sum_{i, j \in n.n. }^{N}{e^2\over |{\vec r}_i-{\vec r}_{j}|}
\end{eqnarray}
where $\vec r_i$ is the location of the $i$-th electron, $i$
and $j$ are restricted to nearest neighbors, and the factor of 
$1/2$ corrects for double counting.  We define the expansion parameter $\epsilon\equiv R/D$
and expand the potential to second order:
\begin{eqnarray}\label{eq:spinU}
 U/E_c &\approx&   U_2 \equiv
  {1\over 2} \sum_{i,j \in n.n.}  \left\{(1+ \epsilon^2) \right. \nonumber  \\
&+& \left. \epsilon^2 \left[{\vec s}_i\cdot {\vec s}_{j}+{3\over 2}
({\hat D_{i,j} }\cdot({\vec s}_i-{\vec s}_{j}))^2\right]\right\} .
\end{eqnarray}
where $E_c \equiv e^2/D$ is a measure of the Coulomb interaction between shells.
Here we specify the position of each charge by a unit vector ${\vec s}_i$ 
pointing from the center of the $i$-th sphere  to the
charge on that sphere. The unit vector ${\hat D_{i,j}}$ 
points from the center of the $i$-th to the center
of the $j$-th sphere.   We have also assumed that the 
lattice possesses reflection symmetry.
From eq.(\ref{eq:spinU}) we see that the system has an AFE  Heisenberg
interaction and a symmetry breaking term.  The symmetry breaking term 
encourages alignment of adjacent spins in the direction of the lattice vector connecting them.  
Further analysis requires knowledge of the lattice vectors of the specific system.

\begin{figure}
  \centering
  {\label{fig:afemodel}\includegraphics[width=2.5in]{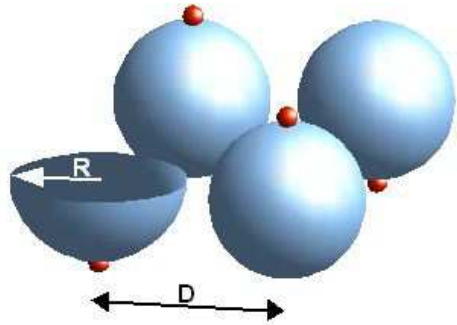}}
\caption{ A schematic picture of the groundstate of classical point
electrons in a 2D array of spherical shells.  Each shell is of
radius R and separated from its neighbor by a distance D.  The 2D
ordering is antiferroelectric for the infinite system size.  
}
  \vspace{-15pt}
  \label{fig:models}
\end{figure}

\subsection{Quantum Mechanical Analysis}

\label{sec:quantum}

Instead of a single point charge on a sphere, in the quantum mechanical
problem the electron is
treated as a charge distribution constrained to the surface of a
spherical shell.  We now have two competing energy
terms: the kinetic, which aims to spread the wave function, and the
potential, which is strictly the Coulomb interaction between nearest
neighbors.  Our starting Hamiltonian is $H=T+U$ where the kinetic 
energy is given by:
\begin{eqnarray}
T&=&-{E_q}\, \sum_i  \, {1 \over \sin{\theta_i}}
{\partial \over \partial \theta_i} {1 \over \sin{\theta_i}} {\partial
\over \partial \theta_i}  + {1\over \sin^2{\theta_i}} {\partial^2
\over \partial \phi_i^2} \\
&=&- {E_q} \sum_i \nabla^2_{\Omega_i} \hfill 
\end{eqnarray}
where $E_q\equiv {\hbar^2 / 2 m^* R^2}$,
and $m^*$ is the effective mass of the electron.
In moving to angular variables, 
we write the unit vectors in eq(\ref{eq:spinU})  in terms of angle variables,
$\vec s_i = \sin{\theta_i} \cos{\phi_i}\, \hat x+  \sin{\theta_i}
\sin{\phi_i} \,
\hat y + 
\cos{\theta_i}  \,\hat z$.   If we measure our energy in units of
$E_q$,
our scaled Schrodinger equation  is then
\begin{eqnarray}\label{eq:tisescaling}
\left(\sum_i \nabla _{\Omega_i}^2 + \lambda \,
U \right)  \psi =\tilde E \, \psi
\end{eqnarray}
where  $\lambda$ is the ratio between the Coloumb  and confinement 
energies, $ \lambda = {E_c / E_q} $ and $\tilde E$ is the 
dimensionless energy.

\begin{figure}
\includegraphics[width=3.25in]{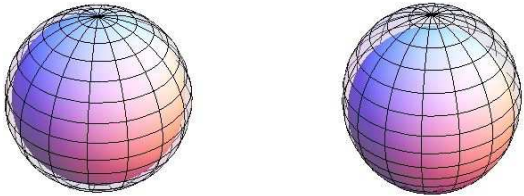}
\caption{Spherical plot showing the charge polarization of a  single lattice site for the 2D square array. Left
is a uniform surface charge density ($1/4\pi$) represented by a mesh sphere of radius $1+ 1/4\pi$ surrounding a
solid sphere of unit radius.  Right is a plot of the polarized charge density, again represented by a spherical
mesh with a radius $1+ \psi^2$ surrounding a solid sphere of unit radius. The charge is not strongly localized
since we are dealing with only the lowest spherical harmonics.
Nearest neighbors display the same behavior but with opposite polarization.}
	\label{fig:chlo}
\end{figure}

We calculate the groundstate 
using a variational approximation. For the $i$-th ring we choose a variational
wave function, using only the first four spherical harmonics ($Y_{\ell m}$):
\begin{eqnarray}\label{eq:vwf} \psi_{\rm var}(\theta_i,\phi_i;\alpha_i;\beta_i,\gamma_i) = 
\cos(\beta_i) Y_{00} + \sin(\beta_i) \cos(\alpha_i) Y_{10} \nonumber \\
+ \sin(\beta_i) \sin(\alpha_i) \sqrt{3 \over 2\pi} \sin(\theta_i) \cos(\phi_i - \gamma_i)
\end{eqnarray}
where we have used time reversal symmetry to force the wave function
to be real.  We are able to use this form since at low energies,
each electron will occupy one of the lowest energy states.  Even at full polarization the charge is not strongly localized; this is demonstrated in 
fig.(\ref{fig:chlo})
If the system will not polarize in this approximation, it will not
be able to do so if we include higher energy states.   This is done for a unit cell that respects the periodicity found in the classical simulations.
Separation between neighbors is assumed
to be sufficiently large to neglect tunneling effects and enable
the use of the Hartree approximation.


\section{2D Square Lattice}

We first analyze the 2D square lattice assuming   periodic boundary conditions.
Below we present the classical and quantum mechanical analyses.

\subsection{Classical Phase Transition}

 The nature of the
symmetry breaking term in the square lattice
can be clarified by resolving the interaction
into components: 
\begin{eqnarray}\label{eq:spinxyzU}
U_2 &= & 
  { 1 \over 2 } \left\{ 
2 \sum_{i}  \left(1 + 2 \epsilon^2 \right) - {3\epsilon^2} ( s_i^{z})^2   \right. \nonumber \\
&+& \left. \sum_{i,j \in n.n.}  
\epsilon^2 \left[-2 {\vec s}_i\cdot {\vec s}_{j}+ 3 s_i^{(z)} s_j^{(z)} \right] \right\}.
\end{eqnarray}  The first term is an irrelevant constant;  the second term breaks
the rotational symmetry, favoring spins aligned along the z-axis.  The third term is 
a Heisenberg {\it ferro}electric coupling while the last and larger term is an Ising
{\it anti}ferroelectric coupling.  An AFE state, when possible,  will maximize the second and
fourth terms, minimizing the total energy.   This is confirmed by Monte Carlo analysis.

\begin{figure} \includegraphics[width=3.25in]{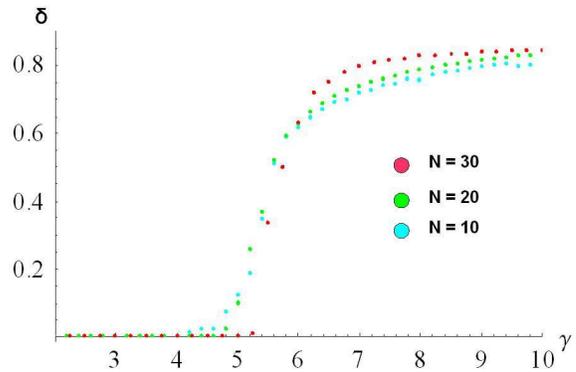} \caption{A
plot of the classical phase transition from random alignment to the
anti-ferroelectric (AFE) state Fig. (\ref{fig:afemodel}) through $\gamma
\equiv \beta E_c$, a measure of interaction strength, at zero
temperature. Here $\delta$ is a measure of the system's AFE order.
Systems are square with periodic boundaries of size N=10, N=20, and
N=30 and have $R/D= 0.417$. Larger system sizes have sharper
transitions.}
\label{fig:ptcl}
\end{figure}

\begin{figure}	
		\includegraphics[width=3.25in]{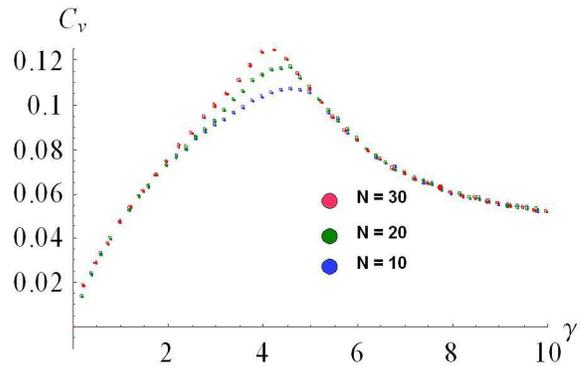}
	\caption{A plot of the heat capacity when $R/D= 0.417$
for N=10, N=20, and N=30 against $\gamma \equiv \beta E_c$, a measure of interaction strength. Here $C_v$ is the system's heat capacity. For larger system sizes, the peak around the transition point is sharper.  }
	\label{fig:hccl}
\end{figure}

Monte Carlo\cite{mc} simulations were performed on 10$\times$10, 20$\times$20 and 30$\times$30 arrays. 
The low temperature ($e^2/kTD)  >> 1$)  configuration 
is  an
antiferroelectric (AFE) pattern, Fig. (\ref{fig:models}).  Furthermore,
a phase transition  from the unordered state to
an AFE aligned state was found numerically.  The transition
is \textit{Ising-like}, with a discontinuous change in the staggered
polarization and a peak in the specific heat which is expected to
sharpen into a divergence as the system size is increased.  
The value of the temperature at the peak extrapolates to $kT_c = XXX$ for an
infinite size system, which should be compared to the expected value 
of $2.24$ for the 2D Ising model.\cite{huang}    

\begin{figure}	
		\includegraphics[width=3.25in]{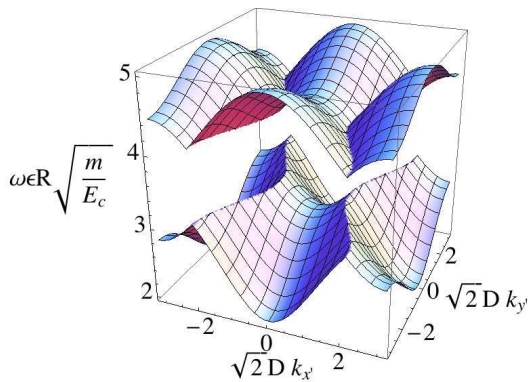}
	\caption{A plot of two of 
the eigenfrequencies of the AFE square lattice as a function of the in
plane wave-vector, $\vec k'$.  The upper surface has been displaced upward
by one unit for clarity.  Without the displacement the two surfaces touch.
The other two normal modes are equivalent to those plotted, but rotated by
$\pi/2$.  Note that all modes are gapped. }
	\label{fig:sqNormModes}
\end{figure}

We can expand about the minimum energy state in order to calculate the 
normal modes. Since the polarization has a checkerboard pattern, our basis
is rotated by $\pi/4$ with respect to the original axis,
and we introduce $\hat x' = (\hat x + \hat y)/\sqrt{2}$ and 
$\hat y' = (\hat x -\hat y)/\sqrt{2}$.
Our unit cell has two elements, an ``up'' site and a ``down'' site.
Denote the position of a point charge on an ``up'' site
by 
\begin{eqnarray}
\vec s_{i}= u_{i}^{(x)}  \, \hat x + u_{i}^{(y)} 
\hat y + \sqrt{1-(u_{i}^{(x)})^2
+(u_{i}^{(y)})^2}\, \hat z
\end{eqnarray}
while that on a ``down'' site is:
\begin{eqnarray}
\vec s_{i}= d_{i}^{(x)}  \, \hat x + d_{i}^{(y)} \hat y - \sqrt{1-(d_{i}^{(x)})^2
+(d_{i}^{(y)})^2} \hat z
\end{eqnarray}
We then solve coupled differential equations
\begin{eqnarray}
m  \,{\partial^2 \vec u_i \over \partial t^2} 
&=&  -{E_c\over R^2} \,  \vec \nabla_u \, U_2 \\
m  \,{\partial^2 \vec d_i \over \partial t^2} 
&=&  -{E_c\over R^2} \,  \vec \nabla_d\,  U_2 
\end{eqnarray}
working to second order in the displacements.  We find find four normal
modes; the first two have frequencies:
\begin{eqnarray}
\omega^{(x')}_\pm = {\epsilon \over R} \sqrt{E_0 \over m} 
\left(10\pm 2 \sqrt{(\cos ({k_y'})-2 \cos ({k_x'}))^2} \right)
\end{eqnarray}
and correspond to in-phase/out-of-phase displacements in the $x'$
direction. These frequencies are  
plotted in figure (\ref{fig:sqNormModes})
as a function of $\vec k'$.
 The second two normal modes (not plotted)
are equivalent to the first but rotated by
$\pi/2$, with $x' \to y'$.   Note that all are gapped.

\subsection{Quantum Phase Transition}

\begin{figure}
\includegraphics[width=3.25in]{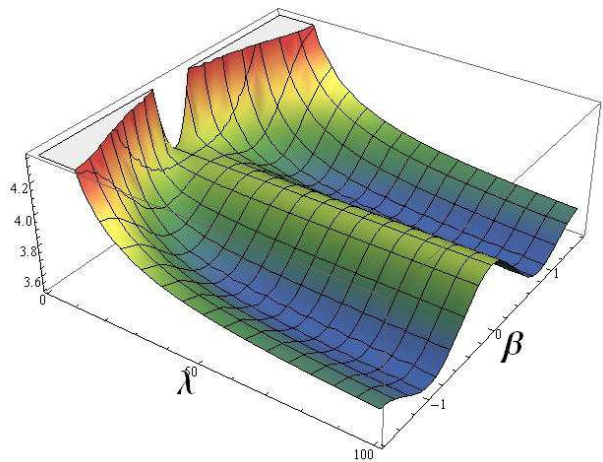} 
\caption{A plot of the
energy functional from the quantum variational calculation for the
2D square lattice. Here, $\epsilon = {1 \over 2}$, thus the ratio
of shell radius to separation is at a maximum (the close-packed
condition). A single minimum for small values of the coupling
constant, $\lambda$, occurs at $\beta = 0$, where $\beta$ is the
mixing parameter found in the variational wavefunction. This state
corresponds with a uniform charge distribution.  As $\lambda$
increases, two minima emerge at $\beta = \pm {\pi \over 4}$. These
values correspond to charge localizing normal to the plane of the
lattice.}
	\label{fig:sqefunc}
\end{figure}

It is not {\it a priori} obvious what will happen in the quantum mechanical
case at zero temperature, since the quantum confinement introduces a new
energy scale.  Rather than solve for the variational 
groundstate of an $N\times N$ system directly, it was assumed that the quantum
groundstate would possess the same symmetry as the classical system, so
that only a $2\times 2$ unit cell of an (assumed) infinite system was
analyzed.  This has the advantage of following the correspondence principle
as $\lambda \to \infty$.  
A plot of the variational energy for 
$\epsilon = {1 \over 2}$, reveals that at small values of the
coupling constant, $\lambda$, a single minimum exists, Fig.
(\ref{fig:sqefunc}). This minimum corresponds to $\beta = 0$, i.e.
a uniform distribution for the electronic wavefunction. As $\lambda$
increases past a critical value, two new minima emerge at $\beta =
\pm {\pi \over 4}$ indicating a quantum phase transition. 
These values correspond to charge localizing
normal to the plane.  Furthermore, nearest neighbors polarize in
opposite directions.

\begin{figure} 
\includegraphics[width=3.25in]{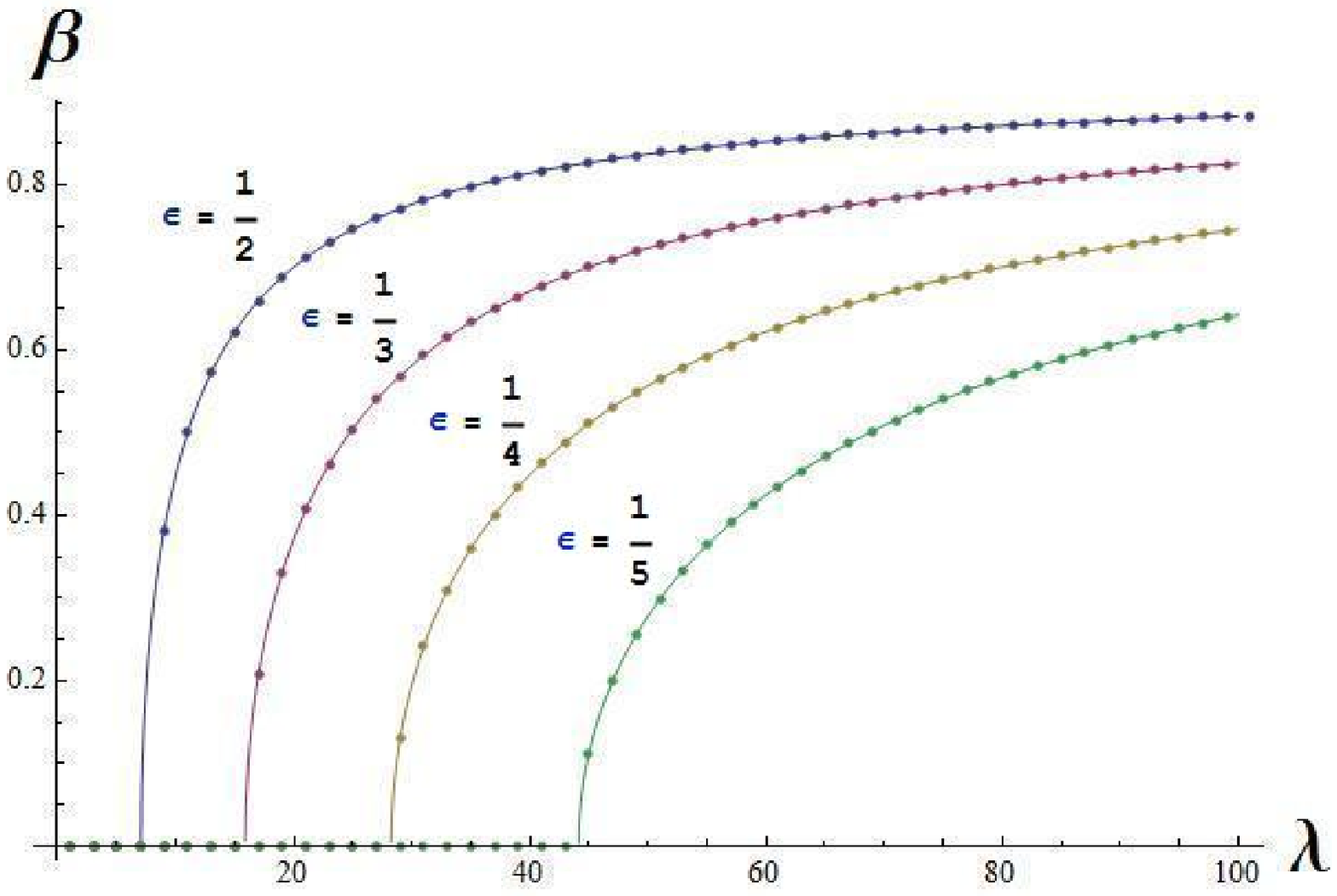} 
\caption{A
plot of the quantum phase transition in the mixing parameter,
$\beta$, of the variational wavefunction through the coupling
constant, $\lambda$,  for several values of
$\epsilon$.   This represents the system's transition from uniform
distribution to antiferroelectric order, through charge localization,
Fig. (\ref{fig:chlo}).  Analytical values of $\beta$ are given by
Eq. \ref{eq:sqbeta}.  Here we see a good fit between numerical and
analytical solutions. } \label{fig:sqtrans} \end{figure}

With the simple form of $\psi_{\rm var}$ of eq.(\ref{eq:vwf}) 
and the approximate
interaction of eq.\ref{eq:spinxyzU}, we can solve 
analytically for the minima of the variational energy and find:
\begin{eqnarray}\label{eq:sqbeta}
\beta = \left\{ 
\begin{array}{l l}
  {1 \over 2} \arccos({192(22 - {175 \over \epsilon^2 \lambda}) \over 35(9\pi^2 - 512)}) & \quad \mbox{if $\lambda \geq \lambda_c$;}\\
  0 & \quad \mbox{if $\lambda < \lambda_c$.}\\ \end{array} \right. \
\end{eqnarray}
where
$\alpha = \pi/2$ and $A = 0$, and the critical value for the
interaction is given by $\lambda_c = {33600 \over (22144 - 315
\pi^2) \epsilon^2}\approx 1.756 /\epsilon^2$.  
Plotting $\beta$ for several values of
$\epsilon$ reveals the phase transition, Fig.
(\ref{fig:sqtrans}).  Here, the analytical minimum  result is
confirmed numerically.

\begin{figure} \includegraphics[width=3.25in]{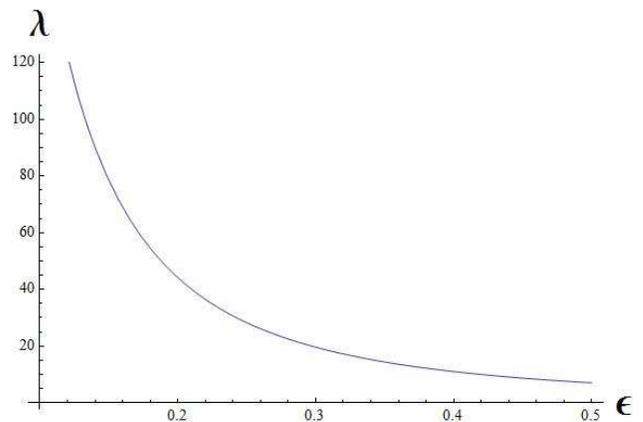} \caption{The
critical value of the coupling constant $\lambda$, 
plotted against the ratio of shell radius to shell separation,
$\epsilon$. It is at these values of $\lambda$ that the transition
in the system occurs, given $\epsilon$.  As separation increases,
the coulomb interaction grows weaker, thus an even weaker confinement
energy is required for the transition to occur.  This corresponds
with a larger physical radius for each shell, $\lambda \gg 1$.}
	\label{fig:sqlam}
\end{figure}

  As $\epsilon$ shrinks, the transition becomes
weaker, requiring the coupling constant, $\lambda$, to be large for
the transition to occur.  This phenomenon emerges from the geometry
of the array.  When the separation between shells becomes larger,
the effective transition radius of each shell also increases, Fig.
(\ref{fig:sqlam}).  The maximum value of $\epsilon$ can have is $1/2$,
when the spheres touch.
When $\epsilon$ is at its maximum there is a minimum $R$ below 
which quantum mechanics dominates and the system does not polarize:
$R_{\rm min} \approx 7 a_0 m/m^*$.
Thus we have the curious result that very small shells do not spontaneously
polarize.


\section{2D Triangular Lattice}

Following our analysis of  the 2D square lattice, we ues  to
the 2D triangular lattice.  Experimental arrays made from the
self-assembly of colloidal particles would likely stack in triangular
arrays, rather than square ones.
For the ring system the triangular 
lattice spontaneously breaks symmetry into a striped AFE phase with a
six-fold degenerate groundstate.
 However, in that case the spins order in the plane, not perpendicular to
it.  

For a pure AFE Ising system the triangular lattice is frustrated.\cite{frustration}
While eq.(\ref{eq:spinU}) is not a simple Ising AFE interaction, it is
clear that frustration will also be present.
Monte Carlo simulation of eqn.(\ref{eq:spinU} )
found multiple low energy states indicative of
a glassy phase.  There was no peak in the specific heat nor any abrupt change in the polarization to indicate an ordered phase.
This meant there was no basis for a quantum mechanical ansatz to the
ground state.

\section{3D Face-Centered Cubic Lattice}

\begin{figure}\label{fig:fcccold}
	\centering
		\includegraphics[width=2.5in]{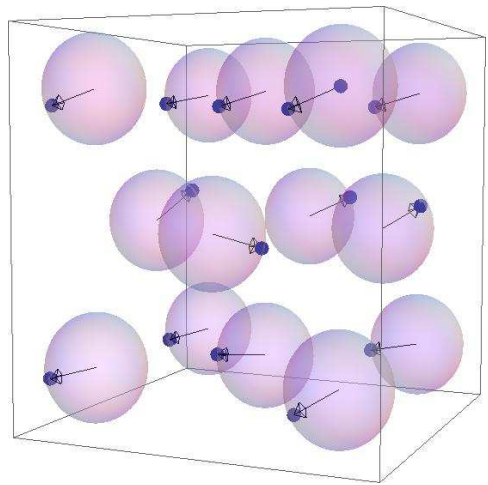}
	\caption{A typical low temperature state that 
 anti-ferroelectric (AFE) in alternating horizontal planes.  Each
	triangular lattice row has a common polarization in the
	plane, while neighboring rows have opposite polarization
	in the plane. }
\end{figure}

The next logical progression is to stack 2D triangular arrays
forming a face-centered cubic (FCC) lattice.  We chose this model since, in the
close-packed limit, we expect it to be more physically reproducible
through fabrication.  Each lattice point has twelve nearest
neighbors: four above, four in the plane, and four below.  Again, the
non-nearest neighbor interaction is considered negligible.

\subsection{Classical Analysis}

\begin{figure}\label{fig:fcccltrans}
\centering
\includegraphics[width=3.25in]{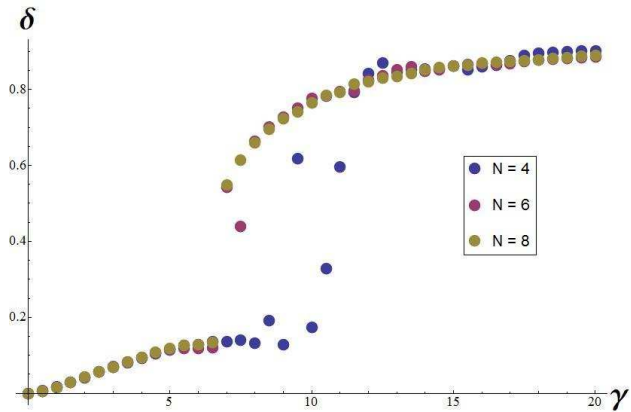}
\caption{A plot of the classical phase transition from random alignment to the anti-ferroelectric (AFE) state Fig. (\ref{fig:fcccold}) through $\gamma \equiv \beta E_c$, a measure of interaction strength, at zero temperature. Here $\delta$ is a measure of the system's AFE order.  Systems are face-centered cubic with periodic boundaries of size N=4, N=6, and N=8. Larger system sizes have sharper transitions.}
\end{figure}

Through numerical Monte Carlo simulation, a minimum energy was shown
to exist when the system is in an AFE state by row. That is, each
triangular lattice row of the FCC structure has opposite polarization,
Fig. (\ref{fig:fcccold}).  Furthermore, a phase transition occurs
from the unordered state to the AFE aligned state.

The temperature of the transition is determined by the interaction
strength found in the Boltzmann factor, $\gamma \equiv \beta E_c =
\beta e^2/D$, where $e$ is the charge of an electron, $\sqrt{2}D$
is the shell separation, and $\beta \equiv 1/kT$.  One can see that
with increased system size comes a sharper transition to the AFE
state, Fig. (\ref{fig:fcccltrans}).  Numerical simulations indicated a
persistent wobble about the in-plane polarization.  Analytic expansion
of the interaction about the uniform unpolarized state indicated an
instability that pushes the polarization slightly off-axis. An expansion about the new minimum indicated restoring forces that were only quartic in the deviation from this state.

\subsection{Quantum Mechanical Analysis}

\begin{figure}
	\centering
		\includegraphics[width=3.25in]{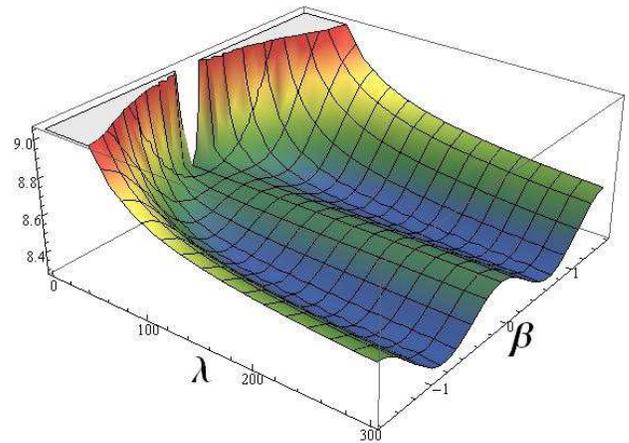}
	\caption{A plot of the energy functional from the quantum variational calculation for the 2D square lattice. Here, $\epsilon = {1 \over \sqrt{2}}$, thus the ratio of shell radius to separation is at a maximum (the close-packed condition). A single minimum for small values of the coupling constant, $\lambda$, occurs at $\beta = 0$, where $\beta$ is the mixing parameter found in the variational wavefunction. This state corresponds with a uniform charge distribution.  As $\lambda$ increases, two minima emerge at $\beta = \pm {\pi \over 4}$. These values correspond to charge localizing along the plane of the lattice.}
	\label{fig:fccefunc}
\end{figure}

\begin{figure}
\centering
\includegraphics[width=3.25in]{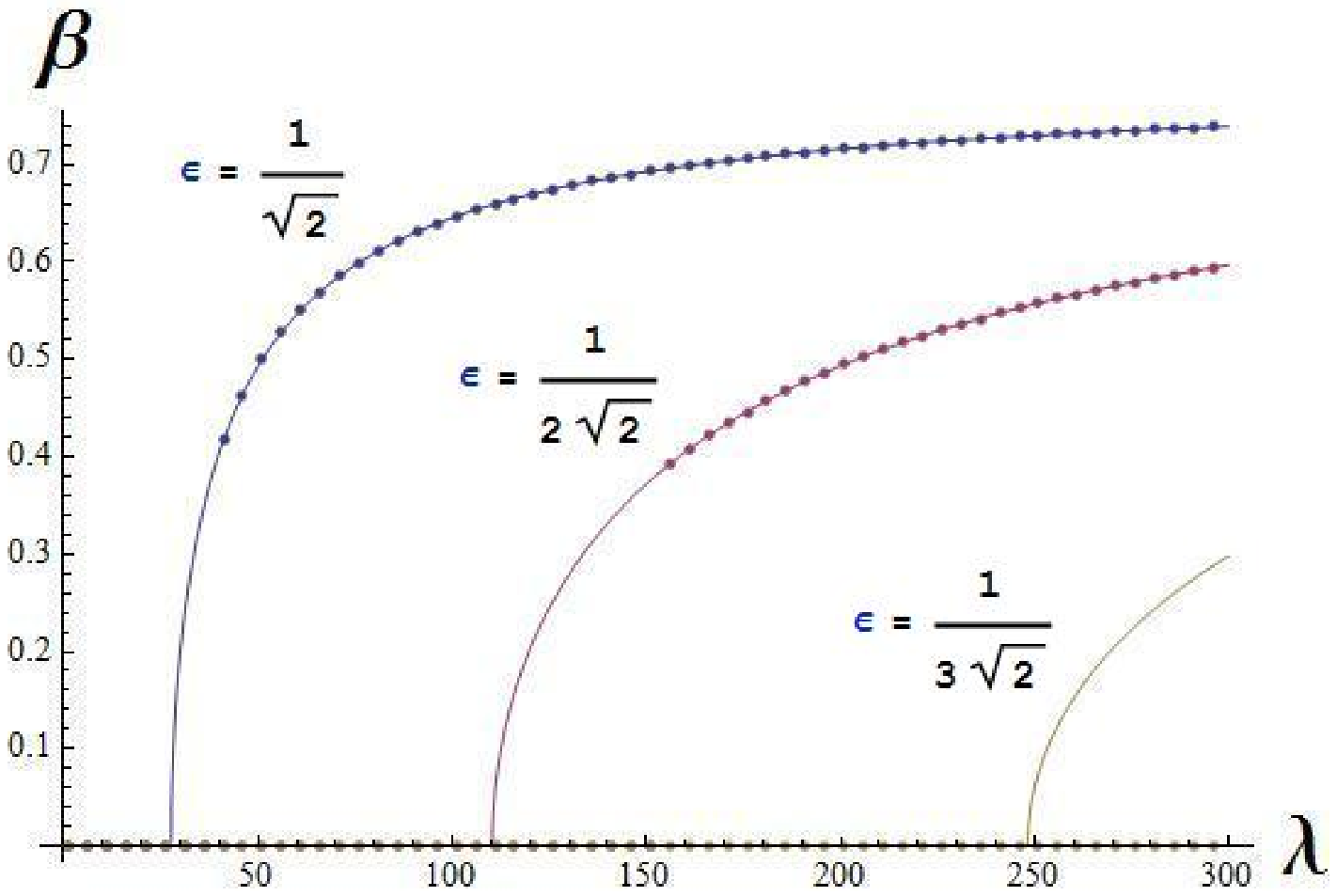}
\caption{
A plot of the quantum phase transition in the mixing parameter,
$\beta$, of the variational wavefunction through the coupling
constant, $\lambda$  for several values of
$\epsilon$.  This represents the system's
transition from uniform distribution to antiferroelectric order,
through charge localization.  Analytical
values of $\beta$ are given by Eq. \ref{eq:fccbeta}.  Here we see
a good fit between numerical and analytical solutions.}
\label{fig:fcctrans} \end{figure}

\begin{figure} \centering \includegraphics[width=3.25in]{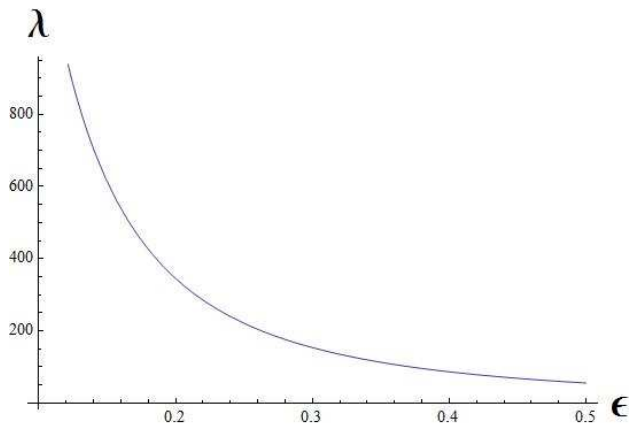}
\caption{ The critical value of the coupling constant $\lambda$,
 plotted against the ratio of shell radius
to shell separation, $\epsilon$. It is at these values of $\lambda$
that the transition in the system occurs, given $\epsilon$.  As
separation increases, the coulomb interaction grows weaker, thus
an even weaker confinement energy is required for the transition
to occur.  This corresponds with a larger physical radius for each
shell, $\lambda \gg 1$.}
	\label{fig:fcclam}
\end{figure}

To confirm our results from classical analysis, we turn to the
Schrodinger variational principle as described earlier.  We assumed
that the  low energy state in
our quantum mechanical computations was polarized in the plane and alternated from one layer to the next, ignoring the slight wobble 
found in the classical analysis.  (This will only give us an upper bound on
the phase transition boundary).  After
completing the variational calculation, we once again find the
charge to localize on each sphere to give the array an anti-ferroelectric
alignment by triangular lattice row, Fig. (\ref{fig:fcccold}).
Furthermore, a quantum phase transition is found from uniform
alignment to this state through the mixing parameter, $\beta$, found
in the variational wavefunction, Fig. (\ref{fig:fcctrans}).

A plot of the energy functional under the close-packed condition,
$\epsilon = {1 \over \sqrt{2}}$, reveals that at small values of
the coupling constant, $\lambda$, a single minimum exists, Fig.
(\ref{fig:fccefunc}). This minimum corresponds to $\beta = 0$, i.e.
a uniform distribution for the electronic wavefunction. As $\lambda$
increases past a critical value, two new minima emerge at $\beta = \pm
\pi/4$. These values correspond with charge localizing along
one of the axis of the shell. Furthermore,  each triangular lattice
row polarizes along opposite axes.

Analytically solving the energy functional for its minima, we find the ground state values of the variational parameters:
\begin{eqnarray}\label{eq:fccbeta}
\beta = \left\{ 
\begin{array}{l l}
  {1 \over 2} \arccos({39 \over \sqrt{2} \epsilon^2 \lambda}) & \quad \mbox{if $\lambda \geq \lambda_c$;}\\
  0 & \quad \mbox{if $\lambda < \lambda_c$.}\\ \end{array} \right. \
\end{eqnarray}
\noindent and where $\alpha = \pi/2$ and $A = \pi/2$, where the critical
value for the interaction is given by $\lambda_c = {39 \over \sqrt{2}
\epsilon^2}$.  Plotting $\beta$ for several values of $\epsilon$ reveals
the aforementioned phase transition, Fig. (\ref{fig:fcctrans}).  Here
again, the analytical minimization result is confirmed numerically.  As
$\epsilon$ shrinks, the transition becomes weaker, requiring the coupling
constant, $\lambda$, to be large for the transition to occur.  This
phenomenon emerges from the geometry of the array.  When the separation
between shells becomes larger, the effective transition radius of each
shell also increases, Fig. (\ref{fig:fcclam}).

It should be noted that this transition is weaker than that discovered for the 2D square lattice.  Not only is the minimum coupling constant larger for each transition to occur, but also $\lambda_c$ increases at a quicker rate with increased separation, Fig. (\ref{fig:fcclam}).

\section{Conclusions}

\label{sec:conclusions}

For a 2D square array and 3D FCC lattice of singly-charge quantum-dot
quantum wells, a phase transition from the unordered state to a
anti-ferroelectric state was found to occur at low energies through
classical Monte Carlo simulation.  Furthermore, this transition is
confirmed through a quantum mechanical treatment of the system in
which only nearest neighbor interactions are considered.  The
transition is a product of the geometry of the system, specifically
dependent upon the ratio of shell radius to shell separation.
Finally, this transition is said to be "Ising-like," which opens
the doors to a host of known collective system behaviors.

For the square lattice we know that in order to have polarization,
$\lambda> 1.75/\epsilon^2$, which means that $R^4m^*/a_0D^3, m> 0.875$,
where
$a_0$ is the Bohr radius.  
For CdSe the effecive mass is $m^*/m=0.13$,
a typical layer thickness is 0.43nm, and the capping (insulating) layer i
about 1.6nm.\cite{pengdots}  If we assume that the shell is two monolayers
thick, then we need shells with a radius of about 7nm in order to have
a polarized ground state.   If the shells have a smaller radius the kinetic
cost to localize the electrons is too high.  Charging the shells may prove
to be too difficult at first; a simpler experiment would be to optically
excite electron-hole pairs in a transverse electric field so as to 
create ``chains'' of elecron-holes across the core-shell array.  
\begin{acknowledgments}

This work is supported by REU grants NSF PHY-0453564, and NSF DMR-0520550.

\end{acknowledgments}

\end{document}